\documentclass[runningheads]{llncs}
\usepackage{graphicx}
\usepackage{pgfplots}
\usepackage{array,multirow,graphicx}
\usepackage{array}
\usepackage{multirow}

\usepackage[utf8]{inputenc}
\usepackage{fourier} 
\usepackage{array}
\usepackage{makecell}
\usepackage{xcolor}

\pgfplotsset{compat=1.14}
\begin{document}
\titlerunning{Fully Connected Layers in Retinal Images}
\title{The relationship between Fully Connected Layers and number of classes for the analysis of retinal images} 
%
%
\author{Ajna Ram \inst{1} 
 \and 
Constantino Carlos Reyes-Aldasoro  \inst{1}}

\authorrunning{A. Ram}
%
\institute{City, University of London, Northampton Square, Clerkenwell, London EC1V 0HB} 
%
\maketitle              
\begin{abstract}
This paper experiments with the number of fully-connected layers in a deep convolutional neural network as applied to the classification of fundus retinal images. The images analysed corresponded to the ODIR 2019 (Peking University International Competition on Ocular Disease Intelligent Recognition)~\cite{ODIR-2019_-_Grand_Challenge}, which included images of various eye diseases (cataract, glaucoma, myopia, diabetic retinopathy, age-related macular degeneration (AMD),  hypertension) as well as normal cases. This work focused on the classification of Normal, Cataract, AMD and Myopia. The feature extraction (convolutional) part of the neural network is kept the same while the feature mapping (linear) part of the network is changed. Different data sets are also explored on these neural nets. Each data set differs from another by the number of classes it has. This paper hence aims to find the relationship between number of classes and number of fully-connected layers. It was found out that the effect of increasing the number of fully-connected layers of a neural networks depends on the type of data set being used. For simple, linearly separable data sets, addition of fully-connected layer is something that should be explored and that could result in better training accuracy, but a direct correlation was not found. However as complexity of the  data set goes up(more overlapping classes), increasing the number of fully-connected layers causes the neural network to stop learning. This phenomenon happens quicker the more complex the data set is. 
\end{abstract}

\section{Introduction}
\subsection{Ocular Diseases}

According to the World Health Organisation (WHO), at least 2.2 billion people have a vision impairment or blindness \cite{WorldHealthOrganisation}. The report stresses that it is {\it at least} 2.2 billion as the number could be much higher. More striking is the fact that from those affected, it is estimated that 1 billion suffer from an impairment that could have been prevented or needs to be addressed. The most common of these cases, and perhaps the simplest to address, is the unaddressed refractive error
(123.7 million) or need of glasses \cite{Foreman_Dirani_Taylor_2017,Honavar_2019}. 

Cataract, or the clouding of the lens of the eye, affects around 65.2 million and can be corrected with surgery and the implant of  intraocular lenses. \cite{Asbell_Dualan_Mindel_Brocks_Ahmad_Epstein_2005}. The surgery is thought to be  the most effective surgical procedure in any field of medicine 
\cite{Thompson_Lakhani_2015}, but still it is not available to certain communities.  

Glaucoma is a disease related to the degeneration of retinal ganglion cells \cite{Weinreb_Aung_Medeiros_2014} and it is estimated that it affects 6.9 million people. 

Short-sightedness, or clinically known as myopia, is a very common eye condition that causes distant objects to appear blurred, while close objects can be seen clearly.

Worldwide, the prevalence of myopia is increasing. Myopia begins at younger ages and progresses faster, leading to more adults with high myopia and risk of sight-threatening complications.~\cite{Wong_Dahlmann-Noor_2020}.
 Myopia is a common cause of vision loss, with uncorrected myopia the leading cause of distance vision impairment globally ~\cite{Holden_Fricke_Wilson_Jong_Naidoo_Sankaridurg_Wong_Naduvilath_Resnikoff_2016}.\\
A high prevalence of age-related macular degeneration, over 14\%, as the cause of blindness and vision impairment, in adults aged 50 years and older in 2015, has been predicted in high-income subregions \cite{Bourne_Stevens_White_Smith_Flaxman_Price_Jonas_Keeffe_Leasher_Naidoo_}.
Function degeneration of macula in aged people has no symptoms in the early stage.

\begin{figure}
\centering
\begin{tabular}{ccc}

\includegraphics[width=35mm,scale=0.5]{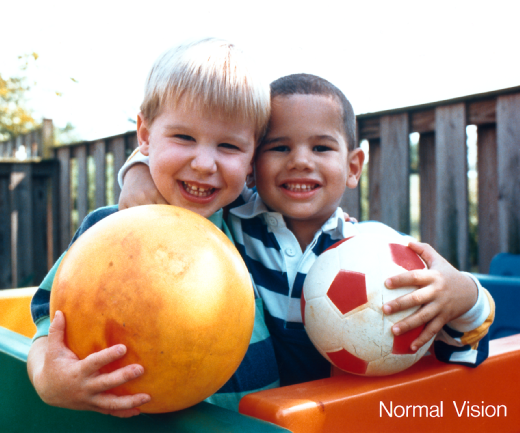}&
\includegraphics[width=35mm,scale=0.5]{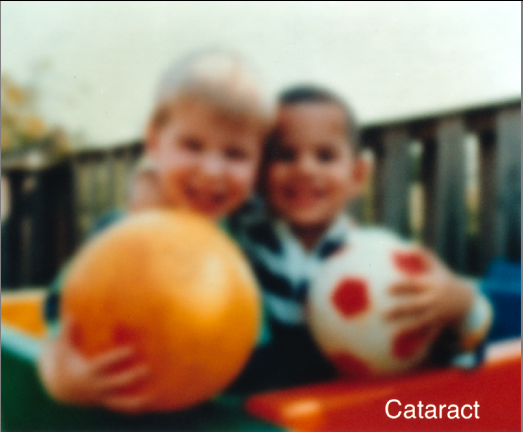} &  \includegraphics[width=35mm,scale=0.5]{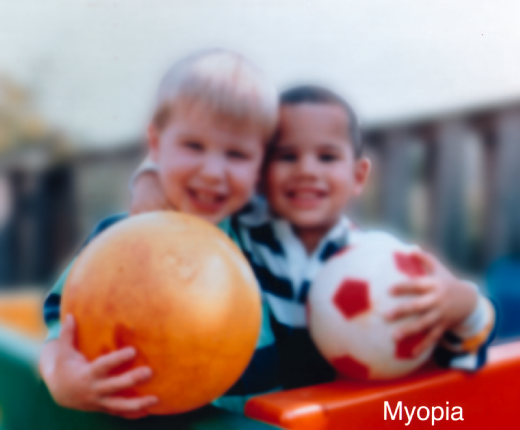} \\ 
\includegraphics[width=35mm,scale=0.5]{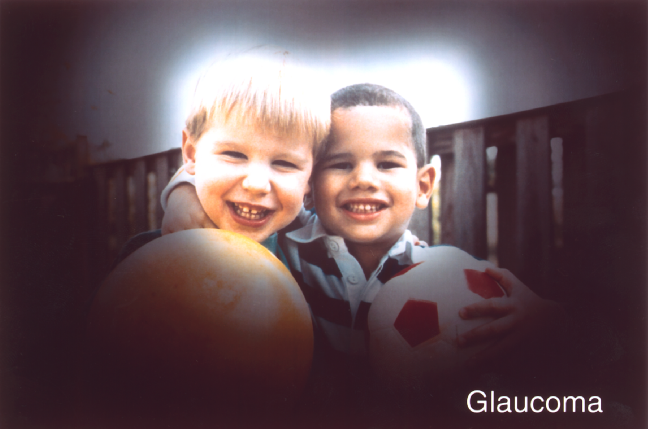} & \includegraphics[width=35mm,scale=0.5]{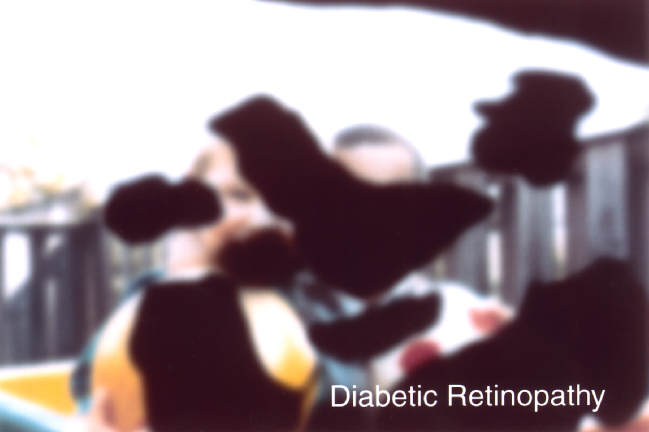} & \includegraphics[width=35mm,scale=0.5]{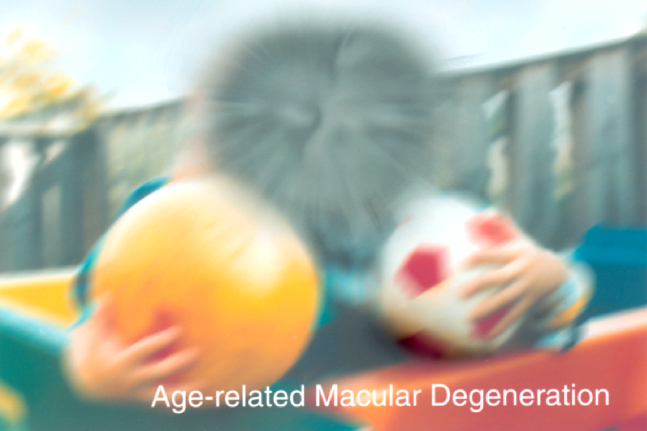} \\
\end{tabular}
\caption{Common eye disease simulations compared with normal vision. Notice how cataracts causes the whole the image to become blurry and cloudy. In glaucoma, the central vision is maintained but the periphery becomes dark. Myopia causes blurriness in vision. Black erratic patches block vision in Diabetic Retinopathy. In AMD, central vision is blocked, while peripheral vision is maintained.  Source~\cite{NEI_NIH}}\label{fig:fig1}
\end{figure}

 Early detection of these diseases could prevent vision damage and other problems. The National Eye Institute, an institute of the National Institute of Health of the USA, has created a series of simulations to illustrate how people with eye disease perceive the world \cite{NEI_NIH}. A sample of those images are shown in Figure \ref{fig:fig1}.

The ocular fundus is the inner lining of the eye made up of the Sensory Retina, the Retinal Pigment Epithelium, Bruch's Membrane, and the Choroid. In ophthalmology, fundus photography is performed for diagnostic purposes; the pupil is dilated with eye drops and a special camera called a fundus camera is used to focus on the fundus. The resulting images can show the optic nerve through which visual "signals" are transmitted to the brain and the retinal blood vessels which supply nutrition and oxygen to the tissue set against the red-orange colour of the pigment epithelium.

Fundus screening allows detection of both ocular and systemic diseases that is diabetes, glaucoma, cataract, age-related macular degeneration (AMD) and other causes~\cite{Fundus_screening_by_medical_technicians}. 
In this paper, data sets consisting of retinal images labelled as normal, cataracts, age-related macular degeneration and myopia are used to understand the effect of increasing the number of fully-connected layers in a Convolutional Neural Network.

\section{Materials}
The images analysed in this paper correspond to the  Pekin University International Competition on Ocular Disease Intelligent Recognition challenge from the Grand Challenge website. This data set contains a structured ophthalmic database of 5,000 patients with age, colour fundus photographs from left and right eyes and doctors' diagnostic keywords were used.

The data set contains patient information collected by Shanggong Medical Technology Co., Ltd. from different hospitals/medical centres in China. In these institutions, fundus images were captured with various cameras in the market, such as Canon, Zeiss and Kowa, resulting into varied image resolutions. There is no patient identifying information. Annotations are labelled by trained human readers with quality control management. 

The images are classified into eight groups including normal (N), cataract (C), AMD (A), myopia(M), diabetes (D), glaucoma (G), hypertension (H), and other diseases / abnormalities (O) based on both eye images and additionally patient age~\cite{NEI_NIH}. For the purpose and intended extensiveness of this paper, only the first four classes were used. An example of a normal fundus image is shown  below in Figure ~\ref{fig7}.
\begin{figure}[!th]
\centering
\includegraphics[width=40mm,scale=0.5]{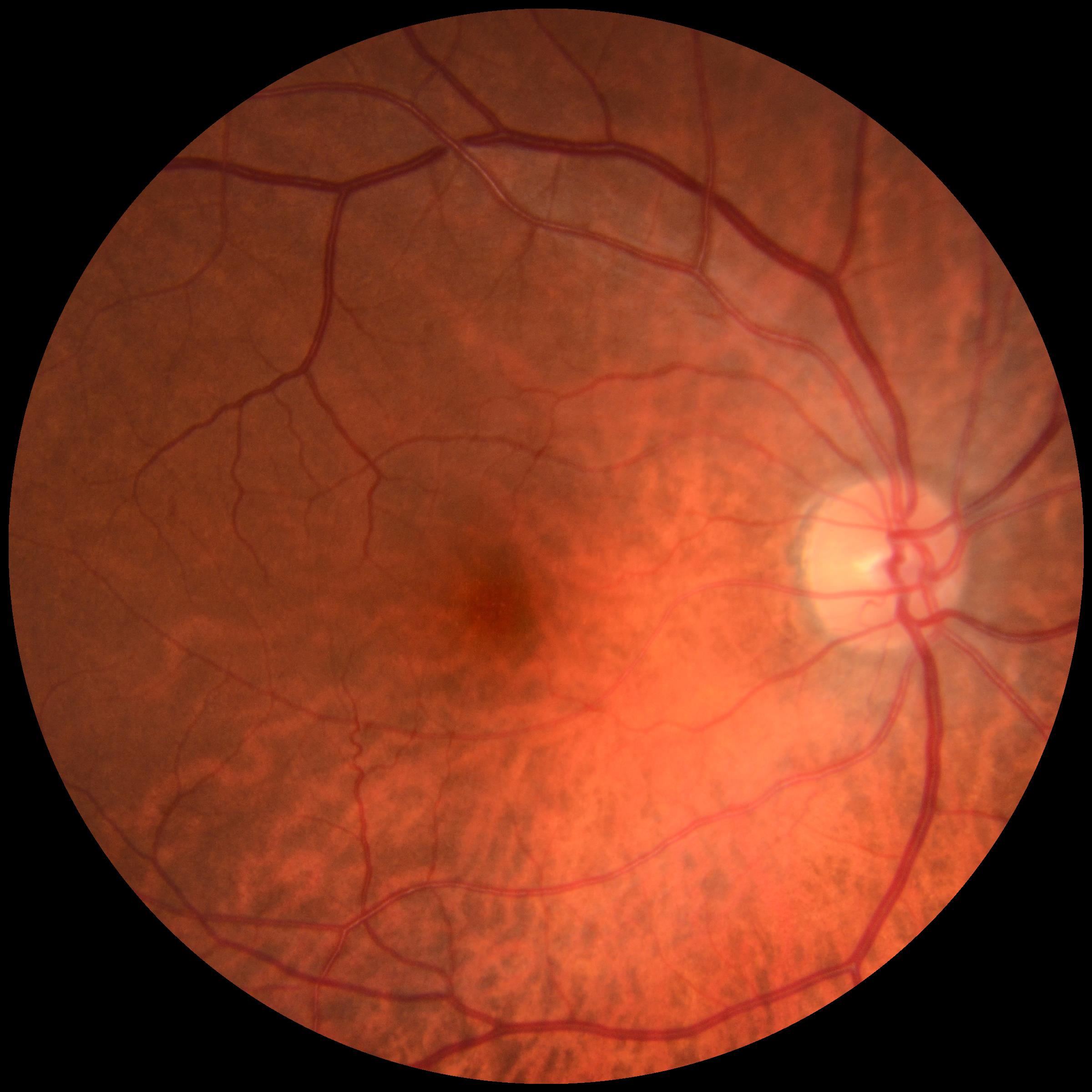}
\caption{Fundus Image of Healthy Patient} \label{fig7}
\end{figure}

\begin{figure}[!ht]
  \centering
\includegraphics[width= 10cm]{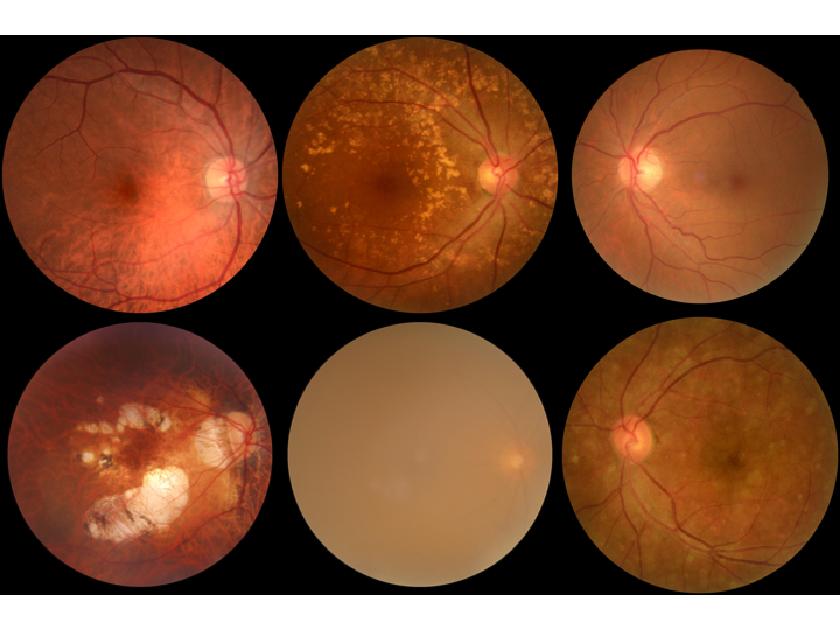} \\
\caption{Six representative images from the data base. These images correspond to the following classes (top left to bottom right): "Normal", "AMD", "Diabetes", "Myopia", "Cataracts", "Other".}
\label{fig:fig2a}
\end{figure}

Three classification comparisons are in this paper, namely:
\begin{itemize}
\item{normal/cataracts (NC)}
\item{normal/cataracts/age-related macular degeneration (NCA)}
\item{normal/cataracts/age-related macular degeneration/myopia (NCAM)}
\end{itemize}
 First case consists of two of the most linearly separable classes; N and C. The classes N, C and A are also easily distinguishable to a trained observed. However, in the third case (NCAM),  age-related macular degeneration (A)  and myopia (M) share similar features which would require more rigorous learning from the CNN.  Six representative images are shown in  Figure ~\ref{fig:fig2a}.

\section{Methods}
\subsection{The role of Fully-Connected Layers in CNNs}
Fully-connected layers, also known as dense layers, connect each neuron in one layer to each of the next layer. The layers map data extracted by previous layers to form the final outcome. The output of convolution/pooling is flatenned into a single vector of values, each representing a probability that a certain feature belongs to a label. 
The output feature maps of the final convolution or pooling layer is typically flattened, i.e., transformed into a one-dimensional array of numbers, and connected to the fully connected layers,  in which every input is connected to every output by a learnable weight. Once the features extracted by the convolution layers and downsampled by
the pooling layers are created, they are mapped by a subset of fully connected layers to the final outputs of the
network, such as the probabilities for each class in classification tasks. The final fully connected layer typically
has the same number of output nodes as the number of classes. Each fully connected layer is followed by a nonlinear function~\cite{FCinCNN}.
\newline
All the code for this work was programmed using Pytorch and executed in Google Co-laboratory "Colab"~\cite{Google_Colaboratory}. 

\begin{figure}[!ht]
\includegraphics[width=\textwidth]{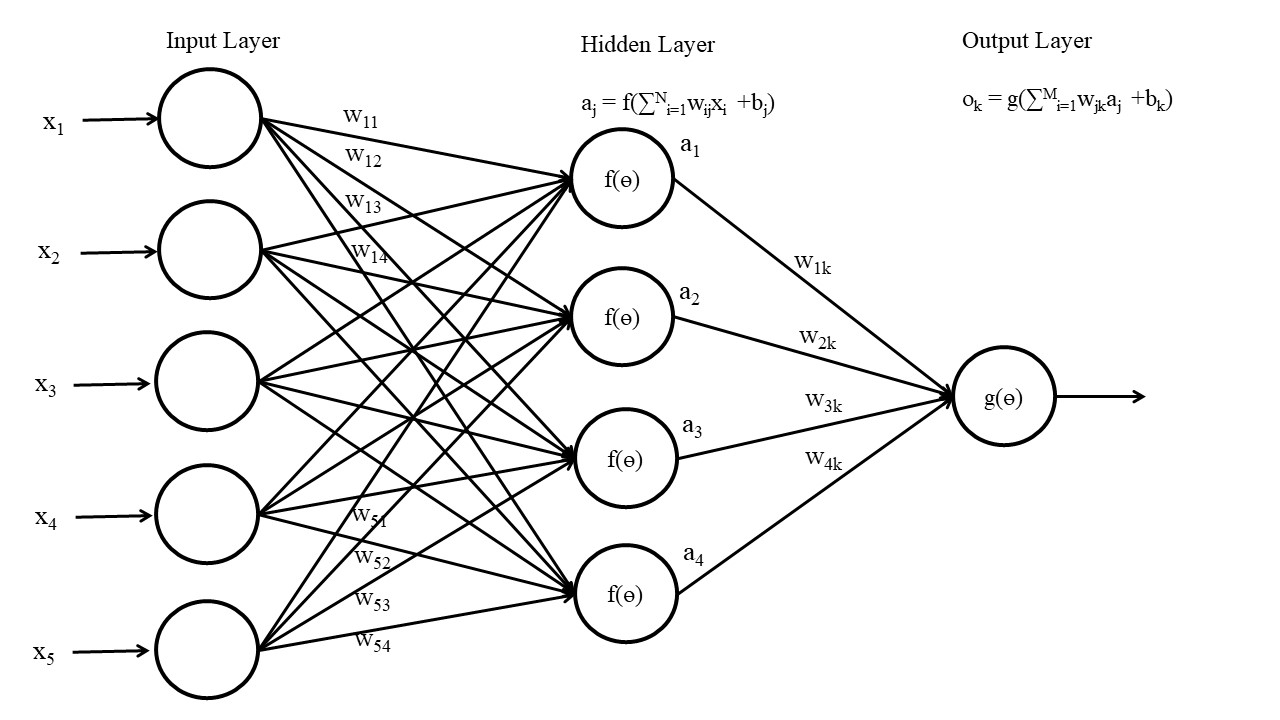}
\caption{Fully-Connected Layers; showing input layer where the feature maps are flattened, hidden Layer, where mapping occurs and output Layer which carries final prediction..}
\label{fig4}
\end{figure}

\subsubsection{Validation of images} Each image was examined for specific image quality factors. These include focus; is focus good enough for grading of smaller retinal lesions, illumination; is image too light or too dark, image field definition; does the primary field include the entire optic nerve head and macula, and artifacts; is the image sufficiently free of artifacts, such as lens dust or light leak from camera.
Images below 1500 x 1500 were discarded, to allow greater dimensional consistency within the data set. 
\subsubsection{Data Pre-processing} A hessian-based filter termed "fibermetric"~\cite{Frangi_Niessen_Vincken_Viergever_1998}, in MATLAB, was applied to the images in order to outline the macula and blood vessels; which are one of the primary elements needed to differentiate one class from another. Figure ~\ref{fig:fig2b} below shows the resulting image after "fibermetric" is applied to it. 
Each data set is balanced by oversampling; such that each class has approximately 1500 images in each.

\begin{figure}[!ht]
  \centering
     \includegraphics[width= 10cm]{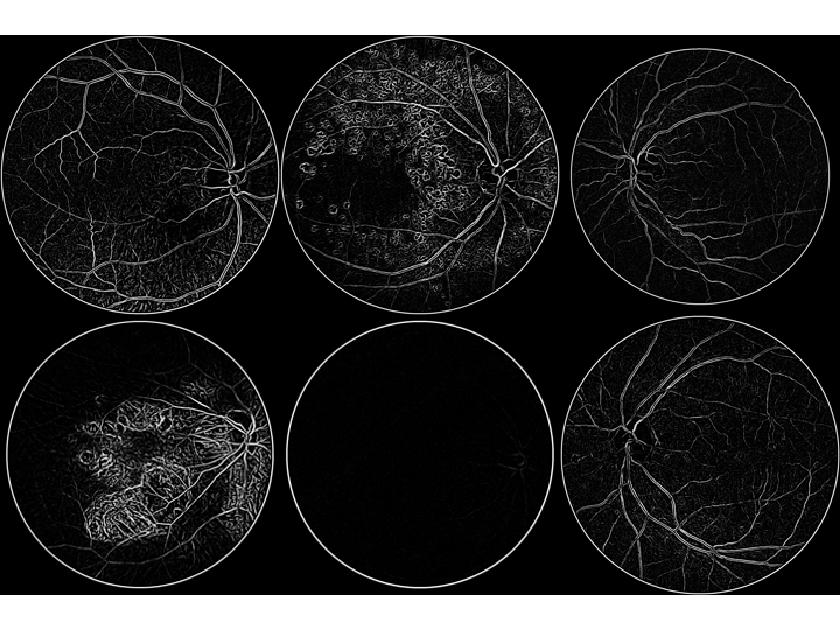}  \\
\caption{Initial processing of the six representative cases shown in Figure~\ref{fig:fig2a}.}
\label{fig:fig2b}
\end{figure}

\subsection{The Convolutional Neural Networks}
The convolutional neural nets employed in this work had a {\it conv-pool-conv-pool} architecture. They have been denoted by CNN-2, CNN-3, CNN-4, CNN-5 and CNN-6, where the numbers represent the number of fully-connected layers in the neural network.  ReLU activation function was applied here. ADAM, with a learning rate of 0.001 was chosen as the optimizer. An overlapping pooling layer with kernel size 5 and stride of 3 was used. 

The number of convolutional layers, their number of input and output kernels and kernel size were kept constant for all CNNs, as shown in Table~\ref{tab1}. Only the number of fully-connected layers and their number of neurons were modified during the experiment, as shown in Table~\ref{tab2}. By keeping the number of convolutional layers constant, the parameters forwarded to fully-connected layers within one data set is the same. And hence, by altering the number of fully-connected, the apparent impression that more layers might lead to better and more specific feature mapping was tested. 

\begin{table}[!ht]
\centering
\caption{Convolutional Architecture of CNNs}\label{tab1}
\begin{tabular}{|c|c|}
\hline
  &Image dimension (2000 x 2000 x 1)\\
\hline
layer-1 &  {\bfseries Conv. (1,6,3), S=1, P=0, ReLU} \\
layer-2 &  {\bfseries MaxPool. (5, S=3)} \\
layer-3 &  {\bfseries Conv. (6,16,5),S=1, P=0, ReLU} \\
layer-4 &  {\bfseries MaxPool. (5, S=3)} \\
\hline
\end{tabular}
\vspace{-7mm}
\end{table}
{\footnotesize *S denotes stride, P denotes padding}

\begin{table}
\centering
\caption{Fully-Connected Layers and their neurons in CNNs}\label{tab2}
\begin{tabular}{|c|c|c|c|c|c|c|}
\hline
FCs & FC-1 & FC-2 & FC-3 & FC-4 & FC-5& FC-6\\
\hline
NN-2 &  {\bfseries(16 x 219 x 219, 128)} & {\bfseries (128 x n)} & {} & {} & {}&{}\\
NN-3 &  {\bfseries(16 x 219 x 219, 96)} & {\bfseries (96 x 8)}& {\bfseries (8 x n)} & {} & {}&{}\\
NN-4 & {\bfseries(16 x 219 x 219, 96)} & {\bfseries (96 x 64)}& {\bfseries (64 x 8)}& {\bfseries (8, n)}& {}&{}\\
NN-5 & {\bfseries(16 x 219 x 219, 96)} & {\bfseries (256 x 128)}& {\bfseries (128 x 64)} & {\bfseries (64 x 8)}& {\bfseries (8 x n)}&{}\\
NN-6 & {\bfseries(16 x 219 x 219, 512)} & {\bfseries (512 x 256)}& {\bfseries (256 x 128)} & {\bfseries (128 x 64)}& {\bfseries (64 x 8)}&{\bfseries (8 x n)}\\
\hline
\end{tabular}
\vspace{-5mm}
\end{table}
{\footnotesize*n denotes number of classes}
\section{Results}

Training accuracy, precision, recall and corresponding F1-score, accuracy per class and the average accuracy of the CNN~\cite{Sokolova_Lapalme_2009} have been considered to assess the performance of the neural network. The data set is shuffled at each run, which introduces some degree of randomness to the results.\\
TP, FP, TN and FN denote true positive, true negative, false positive and false negative respectively. In this work, a TP is considered as an image with a condition, e.g. Glaucoma,  that was classified as Glaucoma. In the same context, a FP is an image incorrectly classified as being glaucoma. A TN is an image correctly identified as not being glaucoma. A FN is a glaucoma picture that has incorrectly been identified as being another class.  
$$ Precision = \frac{TP}{TP+FP}\;\;Recall = \frac{TP}{TP+FN}\;\;F1\;score = \frac{Precision*Recall}{Precision+Recall}$$

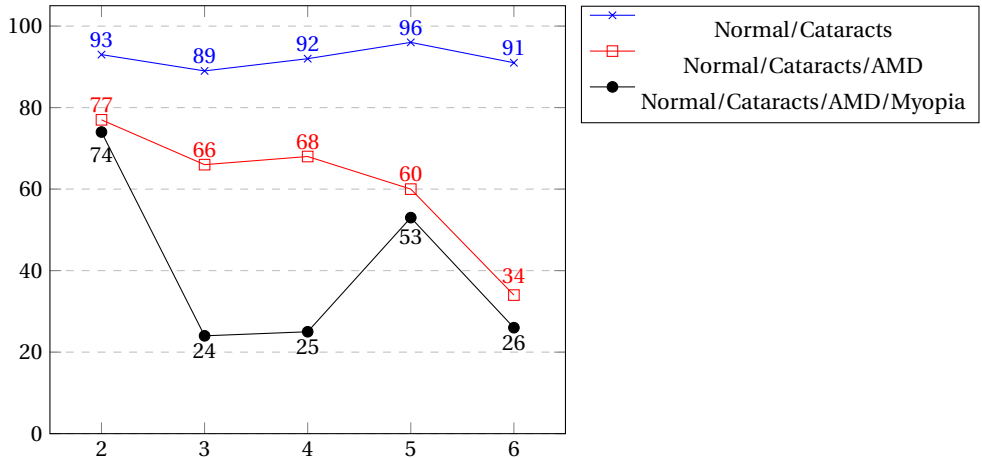
\begin{figure}[!th]
\centering
\begin{tikzpicture}
\begin{axis}[
    xmin=1.5, xmax=6.5,
    ymin=0, ymax=105,
    xtick={1,2,3,4,5,6},
    ytick={0,20,40,60,80,100},
    legend pos=north west,
    ymajorgrids=true,
    grid style=dashed,
    legend style={cells={anchor=north},legend pos=outer north east,}
    ]
\addplot[
    color=blue,
    mark=x,
    ]
    coordinates {
    (2,93)(3,89)(4,92)(5,96)(6,91)
    };
\addplot[ 
    color = red,
    mark = square,
    ]
        coordinates {
    (2,77)(3,66)(4,68)(5,60)(6,34)
    };
\addplot[
    color = black,
    mark = *,
    ]
    coordinates {
    (2,74)(3,24)(4,25)(5,53)(6,26)
    };
    
    \legend{Normal/Cataracts,Normal/Cataracts/AMD,Normal/Cataracts/AMD/Myopia};
\node [above] at (axis cs:  2,  93) {\color{blue} 93};
\node [above] at (axis cs:  3,  89) {\color{blue} 89};
\node [above] at (axis cs:  4,  92) {\color{blue}  92};
\node [above] at (axis cs:  5,  96) {\color{blue} 96};
\node [above] at (axis cs:  6,  91) {\color{blue} 91};
\node [above] at (axis cs:  2,  77) {\color{red} 77};
\node [above] at (axis cs:  3,  66) {\color{red} 66};
\node [above] at (axis cs:  4,  68) {\color{red} 68};
\node [above] at (axis cs:  5,  60) {\color{red} 60};
\node [above] at (axis cs:  6,  35) {\color{red} 34};
\node [below] at (axis cs:  2,  72) {\color{black} 74};
\node [below] at (axis cs:  3,  24) {\color{black} 24};
\node [below] at (axis cs:  4,  25) {\color{black} 25};
\node [below] at (axis cs:  5,  52) {\color{black} 53};
\node [below] at (axis cs:  6,  26) {\color{black} 26};
\end{axis}
\end{tikzpicture}
\caption{Percentage of Training Accuracy against number of FC layers} \label{trainacc}
\end{figure}

{\bf Normal/Cataracts}.
As per Figure ~\ref{trainacc}, training accuracy remain relatively constant, with a standard deviation of 2.3\%, as the number of FCs are varied. 

{\bf Normal/Cataracts/AMD}.
With increase in the number of fully-connected layers, training accuracy plummets, going from 77\% to 34\%. With NN-6, the network predicts a single class. NN-2 got the best average accuracy, out of all the CNNs, with the best F1 score with normal and AMD. NN-3 had the best F1 score for cataracts, however only by a tiny margin, which also proved to have the worse average out of all the CNNS. Refer to Table~\ref{tab10}. 

\begin{table}[!bh]
\centering
\caption{Average training accuracy of CNNs}\label{tab9}
\begin{tabular}{|c|c|}
\hline
  CNN &Avg. Training Accuracy\\
\hline
NN-2 &  {\bfseries 81.3} \\
NN-3 &  {\bfseries 59.7} \\
NN-4 &  {\bfseries 61.7} \\
NN-5 &  {\bfseries 69.7 } \\
NN-6 &  {\bfseries 50.3 } \\
\hline
\end{tabular}
\end{table}

\begin{table}[!ht]
\centering
\caption{Results for Normal/Cataracts/AMD on CNNs}\label{tab10}
\begin{tabular}{|c|c|c|c|c|c|c|c|c|c|}
\hline
\thead {CNN} &\thead {Class} & \thead{Precision} & \thead{Recall} & \thead{F1-score} & \thead{TP}& \thead{TN}& \thead{FP}& \thead{FN}& \thead{Accuracy}\\
\hline
\parbox[t]{2mm}{\multirow{2}{*}{\rotatebox[origin=c]{90}{NN-2}}} & 
Normal & {0.711} & { 0.669 } &  {0.345} & {739}& {1973}&{300}&{366} &{0.803}\\
& Cataracts &  {0.916} &  {0.932} &  {0.462} &  {1037} & {2170} &  {95} & {76} & {0.949}\\
& AMD&  {0.731} &  {0.760} &  {0.373} &  {882} & {1893} &  {325} & {278} & {0.821}\\
\hline
\multicolumn{9}{c}{Average Accuracy of NN-2 for NCA} & {0.858}\\
\hline
\parbox[t]{2mm}{\multirow{2}{*}{\rotatebox[origin=c]{90}{NN-3}}} & 
Normal & {0.514} & { 0.352 } &  {0.209} & {389}& {1905}&{368}&{716} &{0.679}\\
& Cataracts &  {0.913} &  {0.945} &  {0.464} &  {1052} & {2165} &  {100} & {61} & {0.952}\\
& AMD&  {0.541} &  {0.684} &  {0.302} &  {794} & {1543} &  {675} & {366} & {0.692}\\
\hline
\multicolumn{9}{c}{Average Accuracy of NN-3 for NCA} & {0.768}\\
\hline
\parbox[t]{2mm}{\multirow{2}{*}{\rotatebox[origin=c]{90}{NN-4}}} & 
Normal & {0.569} & { 0.462 } &  {0.255} & {440}& {194o}&{333}&{512} &{0.738}\\
& Cataracts &  {0.897} &  {0.932} &  {0.457} &  {1037} & {2146} &  {119} & {76} & {0.942}\\
& AMD&  {0.569} &  {0.710} &  {0.316} &  {824} & {1592} &  {625} & {336} & {0.716}\\
\hline
\multicolumn{9}{c}{Average Accuracy of NN-4 for NCA} & {0.799}\\
\hline
\parbox[t]{2mm}{\multirow{2}{*}{\rotatebox[origin=c]{90}{NN-5}}} & 
Normal & {0.458} & { 0.928 } &  {0.307} & {1025}& {1061}&{1212}&{80} &{0.618}\\
& Cataracts &  {0.915} &  {0.895} &  {0.453} &  {996} & {2173} &  {92} & {117} & {0.938}\\
& AMD&  {0.340} &  {0.254} &  {0.145} &  {18} & {2183} &  {35} & {53} & {0.962}\\
\hline
\multicolumn{9}{c}{Average Accuracy of NN-5 for NCA} & {0.839}\\
\hline
\end{tabular}
\vspace{-2mm}
\end{table}

\begin{table}[!ht]
\centering
\caption{Results for Normal/Cataracts/AMD on CNNs}\label{tab6}
\begin{tabular}{|c|c|c|c|c|c|c|c|c|c|}
\hline
\thead {CNN} &\thead {Class} & \thead{Precision} & \thead{Recall} & \thead{F1-score} & \thead{TP}& \thead{TN}& \thead{FP}& \thead{FN}& \thead{Accuracy}\\
\hline
\parbox[t]{2mm}{\multirow{2}{*}{\rotatebox[origin=c]{90}{NN-2}}} & 
Normal & {0.645} & { 0.581 } &  {0.306} & {642}& {3122}&{353}&{463} &{0.822}\\
& Cataracts &  {0.906} &  {0.919} &  {0.456} &  {1023} & {3361} &  {106} & {90} & {0.957}\\
& AMD &  {0.663} &  {0.714} &  {0.344} &  {911} & {2841} &  {463} & {365} & {0.819}\\
& Myopia &  {0.864} &  {0.861} &  {0.431} &  {935} & {3347} &  {147} & {151} & {0.935}\\
\hline 
\multicolumn{9}{c}{Average Accuracy of NN-2 for NCAM} & {0.883}\\
\hline
\parbox[t]{2mm}{\multirow{2}{*}{\rotatebox[origin=c]{90}{NN-5}}} & 
Normal & {0.382} & { 0.598 } &  {0.233} & {661}& {2404}&{1071}&{444} &{0.669}\\
& Cataracts &  {0.948} &  {0.702} &  {0.403} &  {781} & {3424} &  {43} & {332} & {0.918}\\
& AMD&  {0.437} &  {0.276} &  {0.169} &  {352} & {2851} &  {453} & {924} & {0.699}\\
& Myopia &  {0.527} &  {0.592} &  {0.279} &  {643} & {2918} &  {576} & {443} & {0.778}\\
\hline
\multicolumn{9}{c}{Average Accuracy of NN-5 for NCAM} & {0.766}\\
\hline
\end{tabular}
\vspace{-3mm}
\end{table}
{\footnotesize *Trainings with singular predictions were omitted}

{\bf Normal/Cataracts/AMD/Myopia}. 
Similar behaviour to NCA is observed for this data set. A single class is predicted during training with NN-3, NN-4 and NN-6, hence only ~25\% accuracy is obtained. The best training accuracy achieved for this dataset is with NN-2 with 74\%. Anomalous behaviour is noticed with NN-5 where 53\% accuracy is obtained. \\

NN-2 performance was primarily reduced by not being able to distinguish between Normal and AMD classes. It had more elevated precision, recall and F1 scores for C and M. Refer to Table~\ref{tab6}.\\
NN-5, however had poor performance, with the classes N,A and M with F1 scores of 0.233, 0.169 and 0.279 respectively but did much better with cataracts, which explains its slightly better performance.A precision of 0.948 and recall of 0.702 was obtained for the cataracts class . 

NN-2 performed better overall than the rest of the CNNs.
NN-6 performed the worst out of all, as shown in  Table~\ref{tab9}.

\section{Conclusion} 
\paragraph{}
The results show that for linearly separable data, increasing the number of FCs is potentially a good idea, as the training accuracy does not suffer from it. However, with the amount of data collected it is not possible to predict how the number of FCs is correlated to the accuracy obtained. 

As the number of classes are upped, the performance of the CNNs go down as expected. It can also be observed that as the number of FCs increase, there is a general trend of performance going down. There also seems to be greater occurence of the neural network not learning and making single predictions during training. This shows that mapping of features to a certain class was not done properly with the presence of more layers.
\\
Greater number of FCs therefore does not ensure better feature mapping ability from the network. Increasing it to a certain point causes the network to stop learning altogether and this happens quicker the more complex the data set is.

\section{Further work and Improvements} 
In the context of understanding how FCs affect CNN performance in greater detail, more CNN architectures and more data sets with varying number of classes are to be explored.
\\
Image processing needs further employed to be able to distinguish not so linearly separable classes like AMD and Normal, which posed a problem for the most high performing CNN, that is NN-2.  

%
%

\bibliographystyle{splncs04}
\bibliography{References}

\end{document}